\documentclass[12pt]{article}

\usepackage{array} 
\usepackage{longtable}

\usepackage{jeosman}
\usepackage{amssymb}
\usepackage{amsmath}
\usepackage{subfigure}
\usepackage[latin1]{inputenc}
\usepackage{bm}

\newcommand{\kk}{{\bf k}}
\newcommand{\gb}{{\bf g}}

\begin{document} 

\title{Light propagation in atomic Mott Insulators}
\author{Francesco Bariani}
\address{BEC-CNR-INFM and Dipartimento di Fisica, Universit\`a di Trento,  \\Via Sommarive 14, I-38050 Povo, Italy}
\email{bariani@science.unitn.it}

\author{Iacopo Carusotto}
\address{BEC-CNR-INFM and Dipartimento di Fisica, Universit\`a di Trento,  \\Via Sommarive 14, I-38050 Povo, Italy}


\begin{abstract}
We study radiation-matter interaction in a system of ultracold atoms trapped in an optical lattice in a Mott insulator phase. We develop a fully general quantum model, and we perform calculations for a one-dimensional geometry at normal incidence. Both two- and three-level $\Lambda$ atomic configurations are studied.
The polariton dispersion and the reflectivity spectra are characterized in the different regimes, for both semi-infinite and finite-size geometries.
We apply this model to propose a photon energy lifter experiment: a device which is able to shift the carrier frequency of a slowly travelling wavepacket without affecting the pulse shape nor its coherence.
\end{abstract}

\section{Introduction}

It is a general fact of electrodynamics in continuous media that matter acts as a sort of potential for light:
the dielectric constant in Maxwell equations plays in fact a role mathematically equivalent to the external potential in the Schr\"{o}dinger equation~\cite{russell}.
In particular, remarkable features appear for light propagation through a system with a periodic modulation of the dielectric properties on the scale of the radiation wavelength \cite{yablonovitch, john}: in analogy with electron propagation in crystalline solids, these structures are then called \emph{photonic crystals} (PCs).
According to the Floquet-Bloch theorem, the discrete translational symmetry garantees in fact the conservation of the Bloch wave vector in the first Brillouin zone, the spatial periodicity of the eigenfunctions in the elementary cell, and the separation of the energy spectrum in bands and forbidden gaps~\cite{ashcroft,russell2}.
Depending on the frequency dependence of the dieletric response of the material, PCs are generally classified into two broad categories: passive, if the underlying media are dispersionless, and resonant~\cite{coevorden,erementchouk}.

An interesting example of resonant PC can be realized using ultracold alkali atoms trapped in the periodic potential of an optical lattice~\cite{adams,oberthaler}. In particular, a mile-stone in this field has consisted of the achievement of a Mott Insulator (MI) phase \cite{zoller,greiner}, in which an integer and constant number of atoms are trapped at each site of the periodic optical potential formed by the interference pattern of several laser beams.
If temperature is low enough, atoms are frozen in the lowest vibrational level of each well and the system periodicity is extremely regular: almost no impurities are in fact present (i.e. missing or extra atoms), and there are no phonons.
Thanks to the simple level structure of alkali atoms, one can selectively address specific transitions, so to realize e.g. two or three-level models. From this point of view, the MI can be seen as an extremely resonant PC, where the Bragg scattering processes due to the periodic arrangement of atoms have a strong interplay with the atomic optical resonances. The cleanness of the system, and the weakness of spurious non-radiative effects guarantees that coherence can be preserved for very long times during radiation-matter interaction. 
This is crucial to manipulate the photon propagation without losing its coherence.

In this paper, we develop a general and fully quantum theory to describe the radiation-matter interaction in these systems. Although the model is perfectly general, we specialize our analysis to the 1D case with normally incident light. Both a two-level and a three-level $\Lambda$ configurations are considered.
The band dispersion is characterized as a function of the relative position of the atomic resonance and the Bragg frequency corresponding to the lattice periodicity. Two main regimes are identified depending on whether the two frequency scales are close or well separated. The band dispersion is then used in a calculation of the reflection spectra for both semi-infinite and finite systems. 
These results are the starting point to propose a {\em photon energy lifter} experiment~\cite{zeno}, where the carrier frequency of a slowly travelling wavepacket can be continuously tuned while fully preserving its pulse shape and its coherence.

The paper is organized as follows. 
In Sec.\ref{sec:Hopf}, we review the Hopfield approach to radiation-matter interaction and its application to the case of a lattice of two-level atoms.
In Sec.\ref{sec:1DbandR}, the Hopfield formalism is applied to get predictions for the dispersion of the elementary excitations of the system, the so-called polaritons. Reflection spectra for semi-infinite and finite 1D systems are presented.
In Sec.\ref{sec:1D3level} we extend the model to the case of three-level atoms in a $\Lambda$ level configuration: we implement the dressed atom approach into the Hopfield Hamiltonian and we obtain predictions for the polariton dispersion and reflection spectra.
In Sec.\ref{sec:lifter}, the application of three-level atomic systems as photon energy lifter is discussed for realistic parameters of Rb systems.
The main experimental issues are discussed in Sec.\ref{sec:experiment}. Conclusions are drawn in Sec.\ref{sec:conclusion}.

\section{Hopfield approach}
\label{sec:Hopf}

We consider the interaction between light and a collection of two-level atoms trapped in the periodic potential of an optical lattice \cite{adams,oberthaler}. The atoms are assumed to be in a perfect Mott Insulator state where one and only one atom is present at each lattice site \cite{zoller,greiner}.
With a suitable choice of the lattice laser frequency and polarization, the minima of the optical potential felt by the different atomic states can be made to coincide and to form a cubic lattice of spacing $l$.
Provided trapping at each site is tight enough, the atomic center-of-mass motion results frozen in the motional ground state around the potential minimum (Lamb-Dicke regime) and no atomic tunneling is allowed between adjacent sites.
The energies of the ground $\left|g\right>$ and excited $\left|e\right>$ states of the atom have therefore 
to include the light shift due to the optical potential as well as the zero-point kinetic energy: the resulting energy separation is indicated by $\hbar \bar{\omega}_0$.

Single-atom excitations on the $i^{\textrm{th}}$ atom are respectively destroyed and
created by the operators $\hat{b}_i = |g\rangle_i\,\langle e
|_i$ and $\hat{b}^\dagger_i = |e \rangle_i\langle g |_i$.
In a solid-state terminology, these localized excitations can be seen as an extreme kind of {\em Frenkel excitons} where the electronic excitation is confined to a single atom or molecule of the crystal. This, in contrast to weakly bound Wannier excitons whose electron and hole can be spatially separated by a distance much larger than the lattice spacing \cite{ashcroft}.

In addition to the coupling to the transverse e.m. field that we shall discuss at length in what follows, excitons are coupled to the longitudinal e.m. field. This Coulomb interaction between the dipoles of distinct atoms allows for the transfer of excitation from one site to another according to Hamiltonian terms of the form $\hat{b}^\dagger_i\,\hat{b}_j$  \cite{hopfield} .
To take the most advantage of the translational symmetry of the lattice, one can construct creation operators for delocalized  excitations with a well-defined Bloch wave vector ${\kk}$:
\begin{equation} 
\hat{c}^{\dagger}_{{\kk}} = \sqrt{\frac{1}{N}} \sum_{i}
\hat{b}^{\dagger}_i\, e^{i \kk {\bf l}_i}.
\end{equation}
An integration box of size $L$ is assumed with periodic boundary
conditions. $N=(L/l)^3$ is the total number of atoms in the lattice (which
is assumed to fill the whole integration box). 
The $\hat{c}_\kk,\hat{c}_\kk^\dagger$ exciton operators satisfy the following approximate
Bosonic commutation rules
\begin{equation}
\left[\hat{c}_{{\kk}},\hat{c}^{\dagger}_{{\kk}'}\right] = \delta_{{\kk},{\kk}'} + O\left(\frac{M}{N}\right),
\end{equation}
where $M$ is the number of excitations present in the system. Excitons
therefore behave as bosons, at least in the ``linear optics'' limit
$M\ll N$ in which the probability of a double excitation of the same
atom is negligible. Throughout the whole paper, we shall stick to this
limit.

Thanks to the translational invariance of the system, the Hamiltonian describing the internal atomc dynamics is diagonal in the 
$\hat{c}_{{\kk}}$ operators:
\begin{equation}
H_{at} = \sum_{\kk} \hbar \omega_0(\kk)\,
\hat{c}^{\dagger}_{\kk}\hat{c}_{\kk}.
\end{equation}
The $\kk$ dependence of the exciton band $\omega_0(\kk)$ is a consequence of the Coulomb dipole-dipole interaction and describes the exciton propagation \cite{zoubi}.
As this dependence is quite weak in the present system, we will for simplicity neglect it in the following and take a constant value $\omega_0$ throughout the whole first Brillouin zone (fBz) of the reciprocal lattice. In physical terms, the difference between $\bar{\omega}_0$ and $\omega_0$ corresponds to the Clausius-Mossotti local field correction \cite{hopfield}.

The quantized transverse electromagnetic field is represented by the e.m. vector
potential operator \cite{CCT}
\begin{equation}
\hat{\bf{A}}({\bf x}) = \sum_{{\bf k},{\bf g},\lambda} {\bm
  \varepsilon}_{\lambda} \sqrt{\frac{\hbar}{2 \epsilon_0 L^3
    \omega_{{\bf k}+{\bf g}}}} \left(\hat{a}_{{\bf k}+{\bf g},\lambda} e^{i
  {\bf k x}} + \hat{a}^{\dagger}_{{\bf k}+{\bf g},\lambda} e^{-i {\bf k x}}
\right), 
\end{equation}
where $\hat{a}_{{\kk}+{\gb},\lambda}$ and $\hat{a}^\dagger_{{\kk}+{\gb},\lambda}$ are the
\emph{photon} annihilation and creation operators for the different modes,
labelled by their polarization state ${\bm \varepsilon}_{\lambda=1,2}$
and their wave vector. 
This latter sum is split into the sum over Bloch
wave vectors ${\kk}$ spanning over the fBz of 
the reciprocal lattice, and vectors ${\gb}$ belonging to the
reciprocal lattice. In the cubic lattice geometry under consideration here, the reciprocal lattice is itself cubic, with a lattice constant equal to $2\pi/l$ \cite{ashcroft}.
The free-field Hamiltonian has the usual form 
\begin{equation}
H_{field} = \sum_{{\kk},{\gb}} \hbar
\omega_{{\kk} + {\gb}} \hat{a}^{\dagger}_{{\kk}+{\gb}}\hat{a}_{{\kk}+{\gb}}
\end{equation}
with the vacuum frequency of the photon $\omega_{{\kk}+{\gb}} = c \left|{\kk}+{\gb}\right|$. $c$
is here the speed of light and $\epsilon_0$ the dielectric constant of
vacuum. Throughout the whole paper SI units are in use.

In addition to the terms describing the internal atomic dynamics $H_{at}$ and the non-interacting field $H_{field}$,
the total Hamiltonian has to include terms $H_{int}$ that couple the matter to the transverse e.m. field. This is most simply done by means of the standard minimal coupling replacement \cite{CCT}.
As we are considering optical fields with wavelengths much bigger than
the atomic size, a dipolar approximation can be performed within $H_{int}$. 
Atoms are represented as electric dipoles of dipole moment $\mu$.
For the sake of simplicity, we restrict to a single polarization state
parallel to the dipole moment of the atoms. In this way, we can drop
the polarization index $\lambda$ and write $H_{int}$ in the final,
compact form:
\begin{multline}
H_{int} = \sum_{{\kk},{\gb}}\left[ \frac{ i m }{\sqrt{\omega_{{\kk}+{\gb}}}}\left(\hat{c}_{\kk} \hat{a}^{\dagger}_{{\kk}+{\gb}} + \hat{c}_{-{\kk}} \hat{a}_{{\kk}+{\gb}}\right)+  \textrm{h.c.}\right] + \\ 
+ \sum_{{\kk},{\gb},{\gb}'}\frac{m'}{\sqrt{\omega_{{\kk}+{\gb}}\omega_{{\kk}+{\gb}'}}}(\hat{a}^{\dagger}_{{\kk}+{\gb}} \hat{a}_{{\kk}+{\gb}'} + \hat{a}^{\dagger}_{{\kk}+{\gb}} \hat{a}^{\dagger}_{-{\kk}+{\gb}'} + \textrm{h.c.}).  
\label{eq:hint}
\end{multline}
We can easily recognize the different interacting terms: the former is
the usual dipole exciton-photon coupling, with strength proportional to
\begin{equation}
m= \mu \omega_0 \sqrt{\frac{\hbar}{2\,\epsilon_0\,l^3}},
\label{eq:exphcoupling}
\end{equation}
and the latter is a ``diamagnetic'' photon-photon coupling induced by
the presence of atoms, with a coupling strength proportional to
\begin{equation}
m'= \frac{m^2}{\hbar\omega_0}.
\end{equation}
For typical values of the system parameters, the coupling coefficient
$m$ is generally small, i.e. $m/\sqrt{\omega_0}\ll \hbar \omega_0$,
which implies $m'/\omega_0 \ll m/\sqrt{\omega_0}$.

Thanks to the discrete translational symmetry of the lattice, Bloch
momentum is conserved by all terms in (\ref{eq:hint}). This means that
only states with the same Bloch wave vector are mixed by the
light-matter interaction. 
As this set is discrete, radiative decay (i.e. spontaneous emission)
is forbidden and energy coherently oscillates between the photonic and
the excitonic degrees of freedom \cite{hopfield}. This observation is crucial in
simplifying the physical analysis of the system.

The normal modes of the system, the so-called {\em polaritons} are
therefore superposition of the original photonic and excitonic modes;
they are classified by their Bloch momentum $\kk$, and by a band index
$n$. Thanks to the quadratic structure of the Hamiltonian, the
polaritonic operators can be obtained by means of a Hopfield-Bogoliubov
transformation in the general form~\cite{hopfield}: 
\begin{equation}
\hat{\alpha}_{\kk,n}= x_{\kk,n} \hat{c}_{\kk} + z_{\kk,n} \hat{c}^{\dagger}_{- {\kk}} + \sum_{\gb} \left(w_{\kk,\gb,n} \hat{a}_{{\kk}+{\gb}} + y_{\kk,\gb,n} \hat{a}^{\dagger}_{-{\kk}+{\gb}}\right).
\label{eq:polariton}
\end{equation}
Provided the correct normalization
$|x_{\kk,n}|^2-|z_{\kk,n}|^2+\sum_{\gb}\left(|w_{\kk,\gb,n}|^2-|y_{\kk,\gb,\kk}|^2\right)=1$ is chosen,
the polaritonic operators satisfy bosonic commutation rules
\begin{equation}
\left[\hat{\alpha}_{\kk,n},\hat{\alpha}^\dagger_{\kk',n'}\right]=\delta_{\kk,\kk'}\delta_{n,n'}.
\end{equation}
The coefficients $x,y,w,z$ are determined by solving the eigenvalue
problem associated with the commutator
\begin{equation}
\left[\hat{\alpha}_{\kk,n}, \hat{H}\right] = \hbar \Omega_{\kk,n} \hat{\alpha}_{\kk,n},
\end{equation}
where $\hbar \Omega_{\kk,n}$ is the polariton energy.
The solution of this system is equivalent to the diagonalization of
the following Bogoliubov matrix
\begin{equation}
{\mathcal L}_\kk=
\begin{pmatrix}
{\mathcal K}_\kk & {\mathcal M}_\kk \\
 -{\mathcal M}_\kk^\dagger & -{\mathcal K}_\kk^T
\end{pmatrix}
\label{eq:matrix}
\end{equation}
with the Hermitian matrix 
\begin{equation}
{\mathcal K}_\kk= 
\begin{pmatrix}
\hbar\omega_0  & \frac{im}{\sqrt{\omega_{\kk+\gb}}}& \frac{im}{\sqrt{\omega_{\kk+\gb'}}} & \vdots\\
\frac{-im}{\sqrt{\omega_{\kk+\gb}}} & \hbar \omega_{\kk+\gb} + \frac{2m'}{\omega_{\kk+\gb}} & \frac{2m'}{\sqrt{\omega_{\kk+\gb}\omega_{\kk+\gb'}}} & \vdots\\
\frac{-im}{\sqrt{\omega_{\kk + \gb'}}} &  \frac{2m'}{\sqrt{\omega_{\kk+\gb}\omega_{\kk+\gb'}}}
&\hbar \omega_{\kk+ \gb'} + \frac{2m'}{\omega_{\kk + \gb'}}   & \vdots\\
\cdots & \cdots & \cdots  & \ddots
\end{pmatrix},
\end{equation}
and the symmetric matrix
\begin{equation}
{\mathcal M}_\kk= 
\begin{pmatrix}
0 & \frac{-im}{\sqrt{\omega_{\kk+\gb}}} & \frac{-im}{\sqrt{\omega_{\kk+\gb'}}}  &  \vdots \\
\frac{-im}{\sqrt{\omega_{\kk+\gb}}} & 
 \frac{-2m'}{\omega_{\kk+\gb}} &  
\frac{-2m'}{\sqrt{\omega_{\kk+\gb}\omega_{\kk+\gb'}}} &
\vdots \\
\frac{-im}{\sqrt{\omega_{{\kk}+{\bf g'}}}} &
\frac{-2m'}{\sqrt{\omega_{\kk+\gb}\omega_{\kk+\gb'}}} &
\frac{-2m'}{\omega_{\kk + \gb'}} &
 \vdots \\
\cdots & \cdots & \cdots  & \ddots
\end{pmatrix}.
\end{equation}
The first row and column correspond to the matter excitation. 
For notational simplicity, only two photonic modes of wave vectors
$\kk+\gb$ and $\kk+\gb'$ have been explicitely shown here, but 
the matrices are intended to contain rows and columns for each reciprocal lattice
vector $\gb$.

While the diagonal blocks $ \mathcal{K_{\kk}}$ of $\mathcal{L_{\kk}}$ are hermitian, the non-diagonal $\mathcal{M_{\kk}}$ ones break the hermiticity of the matrix in the usual sense: the matrix $\mathcal{L}_{\kk}$ is in fact ${\bf \Theta}$-hermitian, in the sense that ${\bf \Theta \mathcal{L}_{\kk} \bf \Theta} = \mathcal{L}_{\kk}^{\dagger}$, where ${\bf \Theta} = \textrm{diag}(1,1,1,-1,-1,-1)$ defines the Bogoliubov metric. Physically, this property is related to the fact that the $\mathcal{M}$ blocks correspond to the anti-resonant terms in the Hamiltonian (\ref{eq:hint}), which do not conserve the number of excitations \cite{artoni,ciuti,ciuti2,ciuti3,ciuti4}.
Thanks to the small value of the light-matter interaction coefficients $m$ and $m'$ in the atomic systems under consideration here, most of the physics under investigation here can be obtained by neglecting the anti-resonant terms ${\mathcal M}_\kk$ and  truncating the ${\mathcal K}_\kk$ matrix to a small number of photonic modes~\cite{chong,zoubi}.

\section{1D lattice of two-level atoms}
\label{sec:1DbandR}

\subsection{Band structure}

\begin{figure}[hbtp]
 \begin{center}
 \subfigure[]{\includegraphics[width=0.4\textwidth]{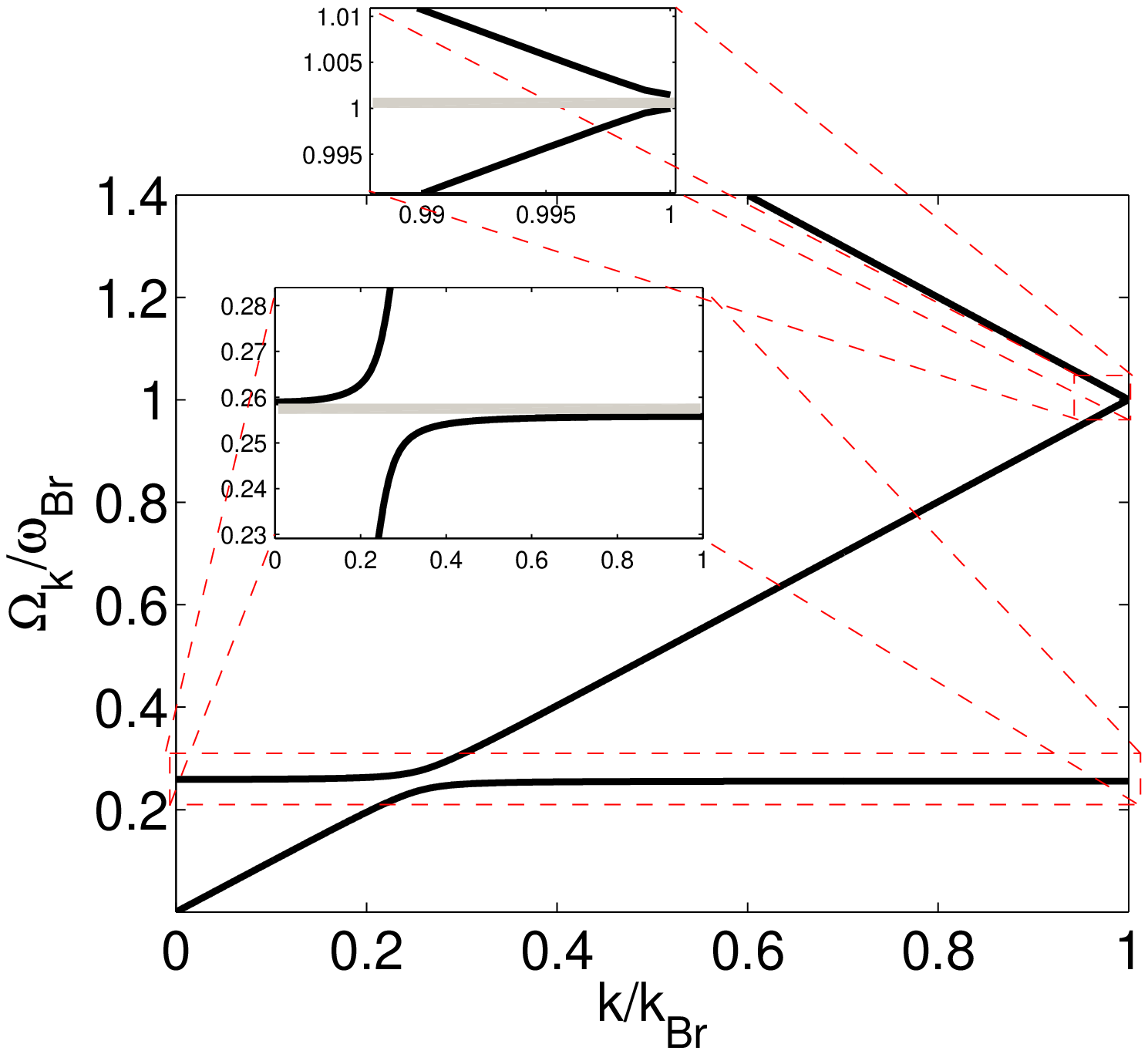} \label{fig:exc}}
 \hspace{0.1\textwidth}
 \subfigure[]{\includegraphics[width=0.4\textwidth]{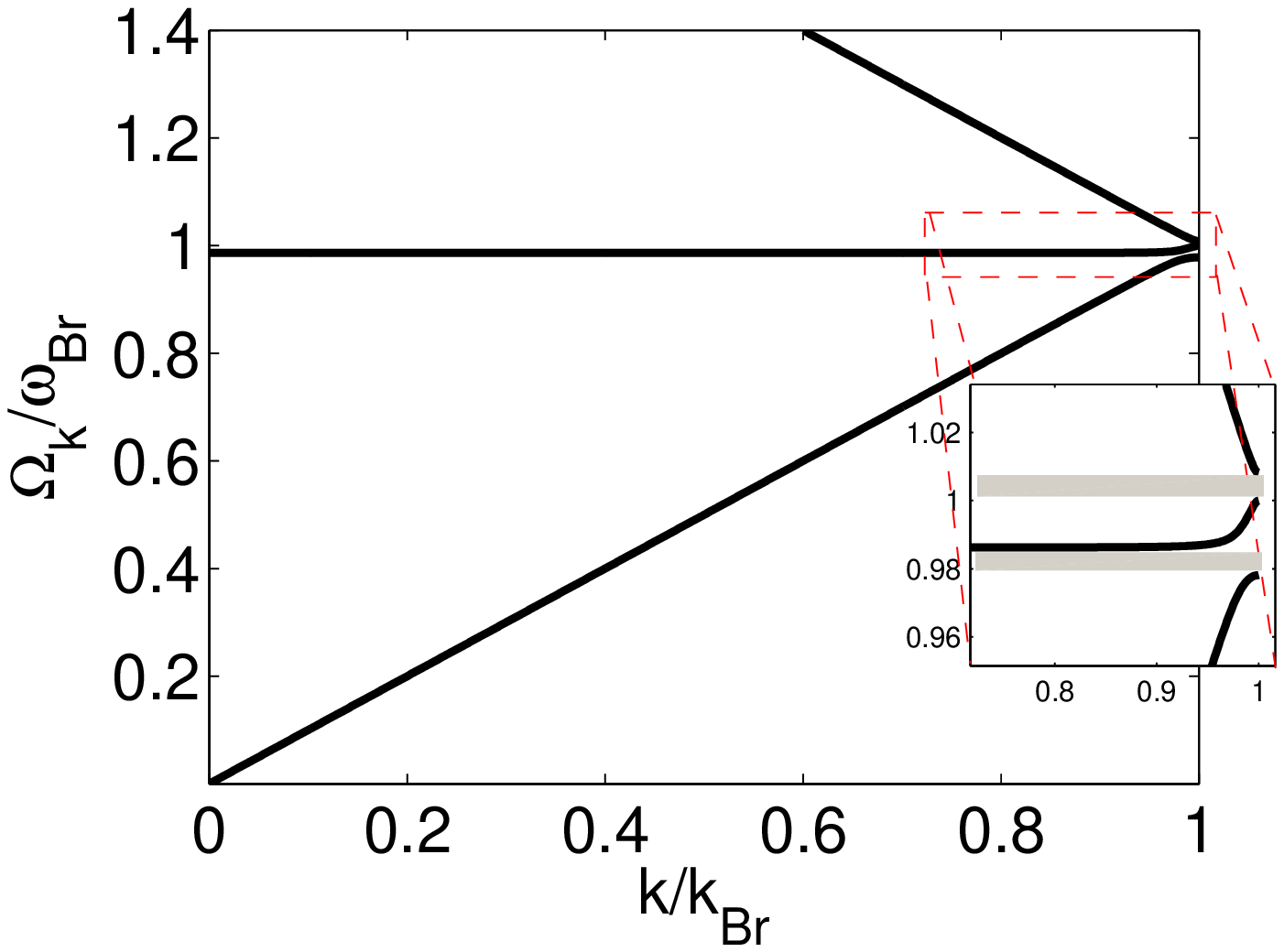} \label{fig:excbragg}} 
\caption{Polariton dispersion in a 1D lattice of two-level atoms. Panel (a): Purely excitonic regime, exciton-photon coupling $(m/\sqrt{\omega_0}) \approx 8 \times 10^{-2} \hbar \omega_0$, distance between resonance and Bragg frequencies $\omega_{Br} - \omega_0 \approx 3\omega_0$. Panel (b): Mixed exciton-Bragg regime, exciton-photon coupling $(m/\sqrt{\omega_0}) \approx 10^{-2} \hbar \omega_0$, distance between resonance and Bragg frequencies $\omega_{Br} - \omega_0 \approx 1.4 \times 10^{-2}\omega_0$. The gray regions correspond to the gaps. For the sake of clarity, the dipole moment has been exaggerated with respect to actual values of atomic systems. }
 \label{fig:disp}
 \end{center}
\end{figure} 
To get a simple physical understanding of the system, it is useful to
concentrate our discussion on the simplest case of a 1D geometry:
most effects related to resonant light-matter interaction are in fact indipendent from the
dimensionality of the system under consideration \cite{coevorden,chong,kempa}.
Interesting discussions of the optical properties of different kinds of 1D resonant PCs can be found in~\cite{ikawa, deutsch, artoni2, artoni3, erementchouk, erementchouk2}.

Two frequency scales are to be considered: the atomic resonance
frequency $\omega_0$, and the Bragg frequency $\omega_{Br}= c\pi/l$ of
the lattice which carries information on the periodicity of the lattice and derives from the famous Bragg condition for diffraction spectroscopy in a reflection geometry \cite{ashcroft}.
The width of the frequency region in which states are effectively
mixed and their splittings are determined by the interacting
Hamiltonian (\ref{eq:hint}). At leading order, this is proportional to
$m/\sqrt{\omega}$.
Starting from this consideration, two different regimes can be
distinguished according to the ratio between the difference
$\hbar(\omega_0-\omega_{Br})$ and the characteristic mixing $m/\sqrt{\omega_{(0,Br)}}$.

The first regime, let's call it {\em purely excitonic} regime
corresponds to the case when the resonance frequency $\omega_0$ and
the Bragg frequency $\omega_{Br}$ are well separated 
$\hbar \left|\omega_{Br}-\omega_0\right| \gg m/\sqrt{\omega_{(0,Br)}}$.
An example of polaritonic dispersion for this regime is shown in
Fig.\ref{fig:exc} for $\omega_0<\omega_{Br}$. For the sake of clarity, a somehow exaggerated value of $m$ has been used in the figure.

On one hand, the exciton and a single photonic mode intersect in the
interior of the fBz; their mixing results in an anticrossing of the
polariton modes with a Rabi splitting equal to
$\hbar \Delta\omega=2 m/\sqrt{\omega_0}$. Far from this region, the
polaritonic modes tend to almost purely photon and exciton modes.
On the other hand, pair of photon modes intersect at the edge of the
fBz at frequencies multiple of $\omega_{Br}$; their mixing is due to
Bragg scattering processes on the atomic lattice.

Gaps open at the edges of the fBz: the lower one (around $\omega_0$)
is the usual polaritonic gap of resonant dielectrics \cite{Jackson}, while the upper
one (just above $\omega_{Br}$) is due to Bragg scattering processes.
Differently from the bulk case, the former gap extends on both sides
of $\omega_0$ because of the limited size of the fBz. The latter one
is instead located strictly above $\omega_{Br}$. Its lower edge is
exactly at $\omega_{Br}$ and corresponds to a purely photonic state,
unaffected by the presence of the atoms which are located at the
electric field nodes. As usual, the polaritonic density of states
vanishes inside the gaps, and radiative propagation at these
frequencies is forbidden.

From a formal point of view, the lower gap mainly originates from the exciton-photon coupling, while the upper one contains a contribution from the direct photon-photon
coupling as well. As $m^2/(\hbar \omega_{Br})\approx m'$, all these terms are
of the same order. This picture is preserved in 2D and 3D, as all
other photonic modes that may participate are far away in energy.

The other regime, let's call it {\em mixed exciton-Bragg} regime, is
characterized by the condition $\hbar(\omega_{Br}-\omega_0) \lesssim
m/\sqrt{\omega_{(0,Br)}}$. 
In this case three modes are simultaneously mixed, namely two photon
branches (the incoming one at $k$ and the first Bragg diffracted at
$k-2\pi/l$) and the excitonic state. 
Differently from the previous case, the splittings (again of the order
of $m/\sqrt{\omega_{(0,Br)}}$) are now located close to the edges of the
  fBz. As one can see in Fig. \ref{fig:excbragg}, this results in much wider
forbidden gaps of the order of the splitting. 
In addition to this, the ``central'' band between $\omega_0$ and
$\omega_{Br}$ results squeezed and shows a very flat dispersion over
most of the fBz.

In the mixing region around the band edge, the upper and lower
polaritons are mixture of exciton and photons with almost equal
weights, while the central polariton band at $\omega_{Br}$ is mostly
photonic. This effect is easily explained in terms of the electric
field showing nodes at the atomic locations, as in the previous regime.

\subsection{Reflectivity spectra}

The band dispersion introduced in the previous section is a complete
description of the photon propagation inside the lattice.
Most spectroscopic experiments, however, involve light beams which are
incident onto finite systems and therefore require a description of
the interfaces between regions of different optical properties, namely
the external vacuum and the atomic lattice. This allows to calculate
crucial properties of the system, such as its reflectivity~\cite{erementchouk,artoni3, artoni2, deutsch, ikawa}.

This can be done by imposing suitable boundary conditions on the
electromagnetic fields: Maxwell theory imposes in fact the
continuity of both the electric field and its spatial derivative.
In the following, we shall restrict to the most relevant case of a
coherent, monochromatic excitation at $\omega$. Outside the lattice,
we have purely photonic, coherent plane waves with wave vectors such that
$|k|=\omega/c$. Inside the lattice, the field propagates as coherent
polaritonic states of modes such that $\Omega_{k,n}=\omega$.

The electric field is the expectation value of the electric field
operator
\begin{equation}
\hat{E}(x) = i \sqrt{\frac{\hbar}{2\epsilon_0 L^3}} \sum_{q}
\sqrt{\omega_{q}} \left(\hat{a}_q e^{iqx} - \hat{a}^{\dagger}_q e^{-iqx}\right).
\label{eq:efieldoperator}
\end{equation}
Outside the lattice, the electric field associated with a photonic
coherent state of wave vector $k$ is given by   
\begin{equation}
E(x,t) = \left<coh:a_k(t)\right|\hat{E}(x)\left|coh:a_k(t)\right> = i
  \sqrt{\frac{\hbar \omega_k}{2\epsilon_0 L^3}}\,a_k\,e^{i(kx - 
  \omega_{k}t)} + \textrm{h.c.}.
\label{eq:E_out}
\end{equation} 
Inside the lattice, one is to consider a polaritonic coherent state with Bloch wave vector $k$ in the $n$ band 
satisfying
\begin{equation}
\hat{\alpha}_{k,n}\,|coh:\alpha_{k,n}(t)\rangle=\alpha_{k,n}(t)\,|coh:\alpha_{k,n}(t)\rangle;
\end{equation}
time evolution of such a state under the system Hamiltonian corresponds to 
\begin{equation}
\alpha_{k,n}(t)={\alpha}_{k,n}\,e^{-i\Omega_{k,n} t}={\alpha}_{k,n}\,e^{-i\omega t}.
\end{equation}
The electric field of such a state is then:
\begin{multline}
E(x,t)=\left<coh:\alpha_{k,n}(t)
\right|\hat{E}(x)\left|coh:\alpha_{k,n}(t)\right> = \\ 
= \left[ \sum_g
  i \sqrt{\frac{\hbar\,\omega_{k+g}}{2\epsilon_0 L^3}}\, {\alpha}_{k,n}\, (w^*_{k,g,n} + y^*_{k,-g,n})\, e^{i((k+g)x -
    \Omega_{k,n}t)} + \textrm{h.c.}\right]. 
\label{eq:polaritonfield}
\end{multline}
The amplitudes $a_k$ and $\alpha_{k,n}$ are then determined by matching the fields (\ref{eq:E_out}) and (\ref{eq:polaritonfield}) at the interfaces for the given geometry under consideration.

\begin{figure}[hbtp]
 \begin{center}
 \subfigure[]{\includegraphics[width=0.4\textwidth]{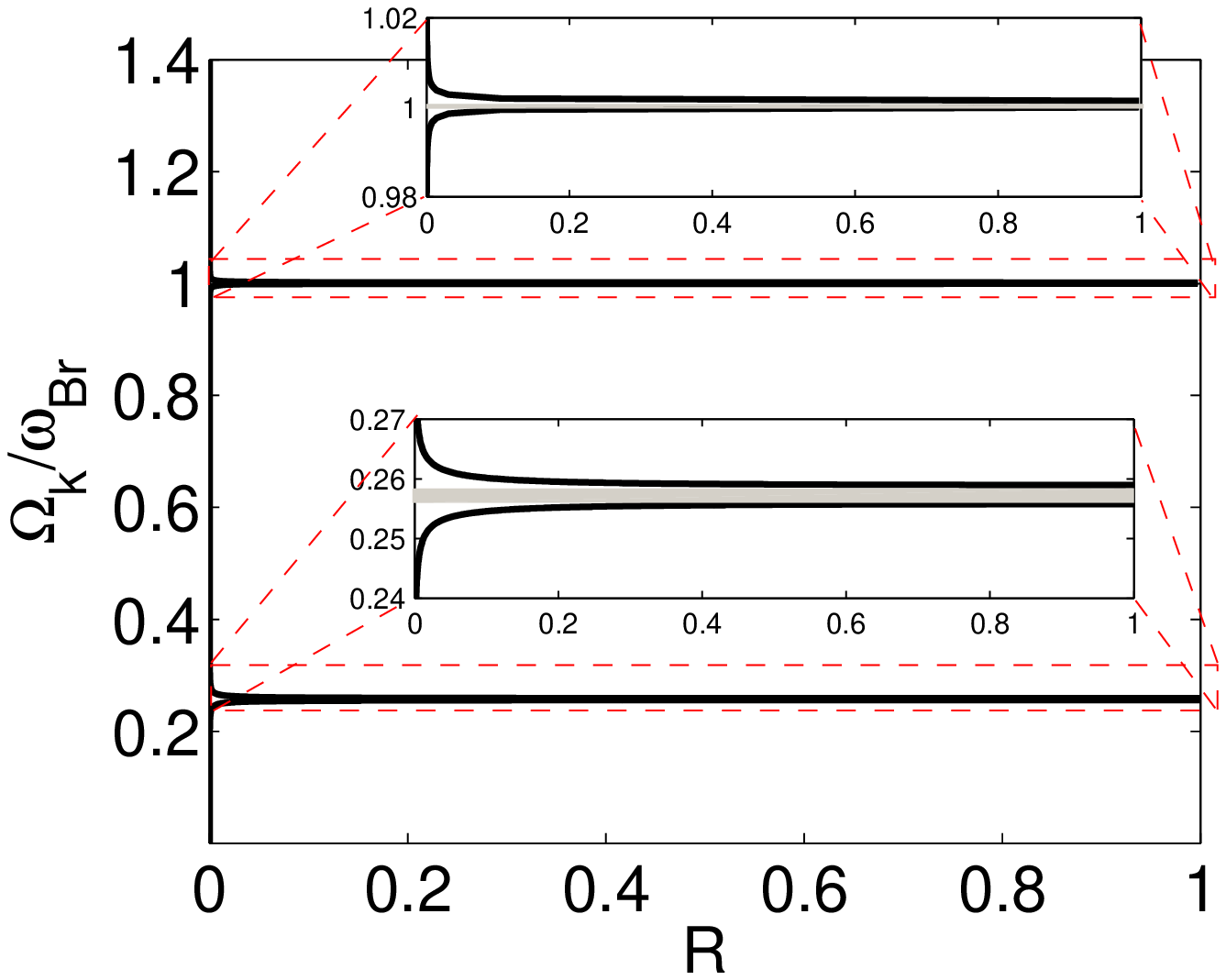} \label{fig:Rsemiinfexc}}
 \hspace{0.1\textwidth}
 \subfigure[]{\includegraphics[width=0.4\textwidth]{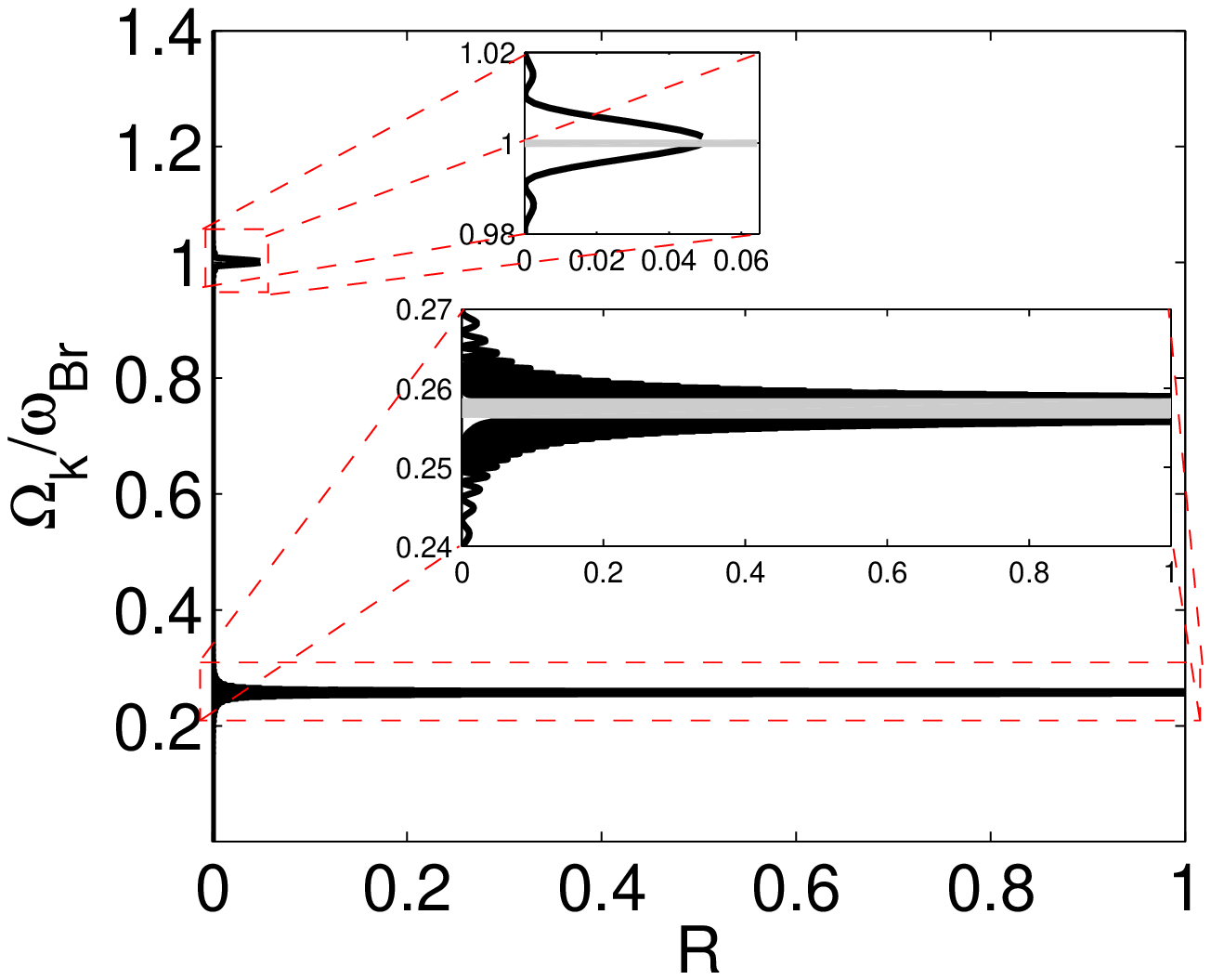} \label{fig:Rfinexc}}
\caption{Reflectivity spectra in the purely excitonic regime for (a) a semi-infinite lattice, (b) a finite one containing $M = 100$ cells. Parameters as in Fig. \ref{fig:disp}. The gray regions correspond to the gaps.}
  \label{fig:Rexc}
 \end{center}
\end{figure}

\begin{figure}[hbtp]
 \begin{center}
 \subfigure[]{\includegraphics[width=0.4\textwidth]{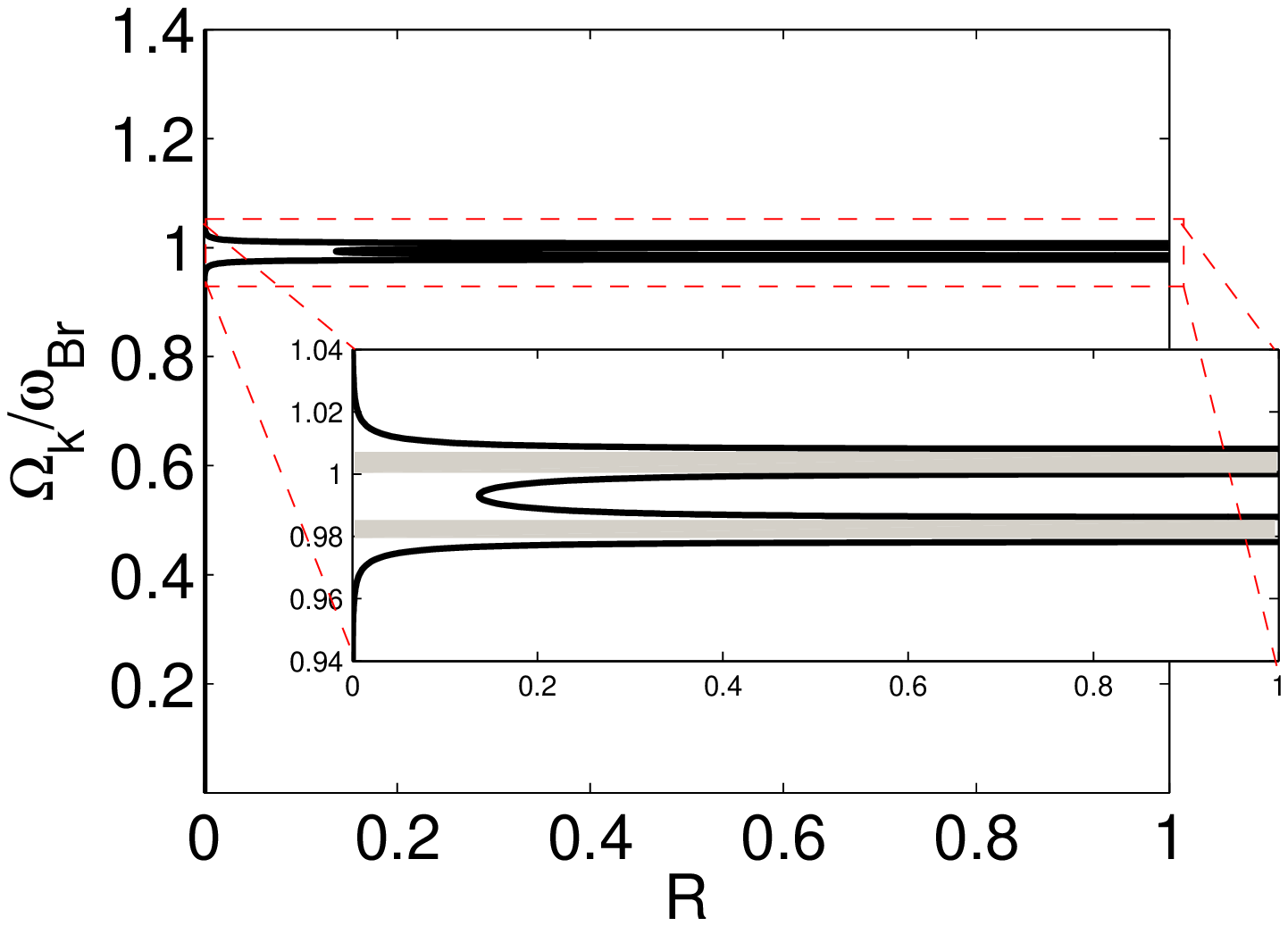} \label{fig:Rsemiinfexcbragg}}
 \hspace{0.1\textwidth}
 \subfigure[]{\includegraphics[width=0.4\textwidth]{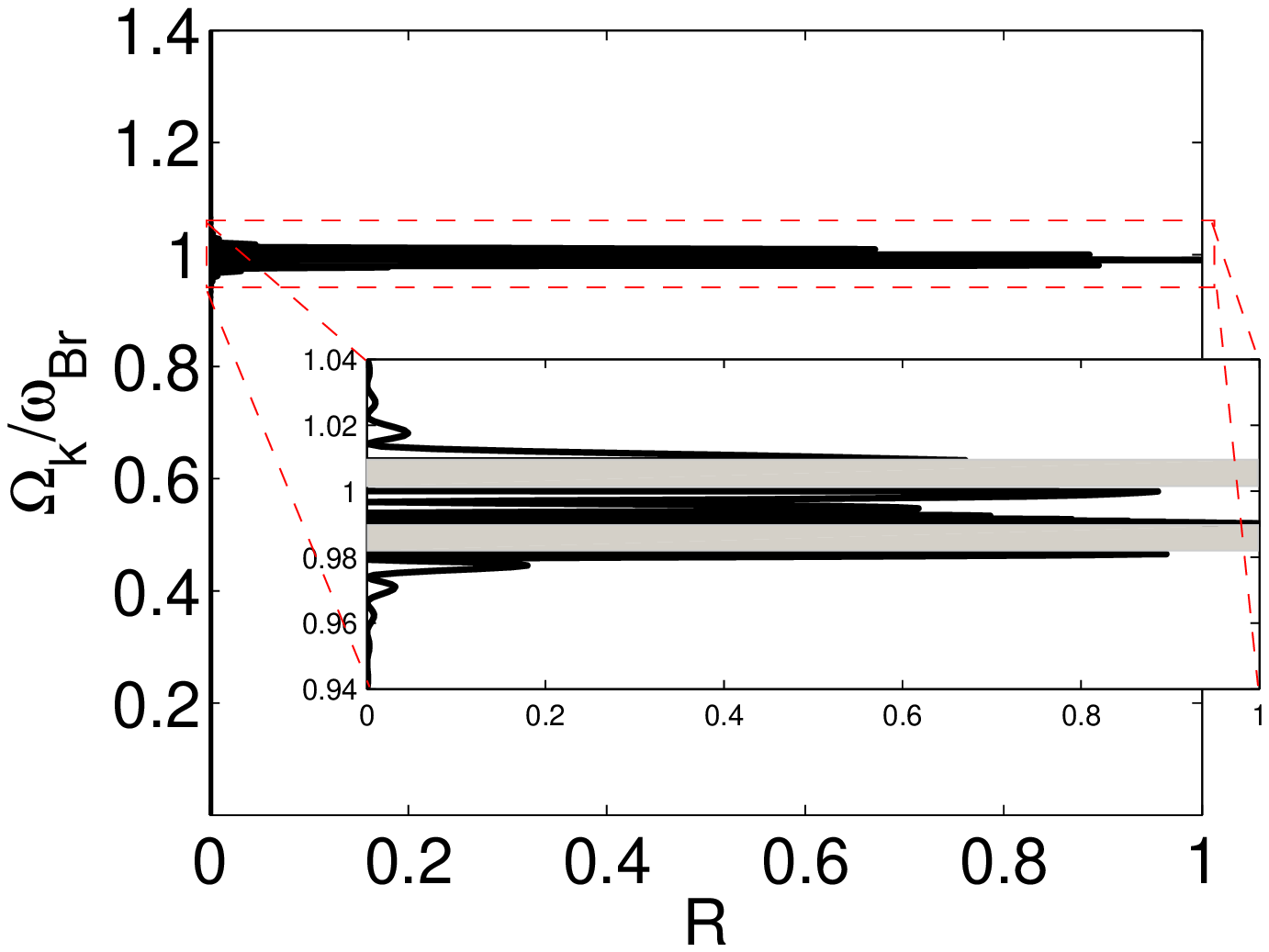} \label{fig:Rfinexcbragg}}
\caption{Reflectivity spectra in the mixed exciton-Bragg regime for (a) a semi-infinite lattice, (b) a finite one containing $M = 100$ cells. Parameters as in Fig. \ref{fig:disp}. The gray regions correspond to the gaps.}
 \label{fig:Rexcbragg}
 \end{center}
\end{figure}

The plots in Fig. \ref{fig:Rexc} and Fig. \ref{fig:Rexcbragg} show the reflectivity
spectra in respectively the purely excitonic and the mixed exciton-Bragg regimes.
Two geometries will be considered: a semi-infinite [panels (a)], and a finite one [panels (b)]. 
Note that these predictions exactly coincide with the ones of semi-classical calculations where matter is described in terms of a dielectric polarizability \cite{coevorden,deutsch,kempa,erementchouk,erementchouk2,artoni2,artoni3}.

\subsubsection{Semi-infinite geometry}

In the first case, Fig. \ref{fig:Rsemiinfexc} and Fig. \ref{fig:Rsemiinfexcbragg}, there is a single interface, dividing the space in two semi-infinite regions: vacuum and lattice. 

We first consider the {\em input} problem with an incoming and a reflected plane wave in the vacuum and a single transmitted polariton Bloch mode in the lattice: the Bloch momentum $k$ in the fBz and the band index $n$ of this latter are chosen in order to satisfy energy conservation $\omega=\Omega_{k,n}$. The reflection amplitude $r_{in}$ is then expressed in terms of the electric field of the polariton and its spatial derivative as
\begin{equation}
r_{in} = e^{2i(\omega/c)(l/2)}\frac{E(l/2)+i(c/\omega)E'(l/2)}{E(l/2)-i(c/\omega)E'(l/2)}.
\label{eq:Rin}
\end{equation}
The point $x=l/2$ lies on the edge of the elementary cell, i.e. midway between neighbouring atoms
\footnote{Actual numerical calculations are performed by truncating the full matrix (\ref{eq:matrix}) to a finite number of modes; in order to obtain a smooth convergence, a gaussian cut-off has also been added on higher photonic modes. This cut-off physically mimicks finite size atoms.
Because of the singularity in the field at the atomic position $x=0$, evaluating the fields at $x=l/2$ rather than at $x=0$ ensures faster convergence.
}.

The mismatch between the photonic components of the polariton state and the incoming wave determines the reflectivity $R_{in}=|r_{in}|^2$ shown in the figures: this is significant around the gaps where the incoming wave is strongly mixed with either an exciton [around the lower gap in Fig.\ref{fig:Rsemiinfexc}], a photon [around the upper gap in Fig.\ref{fig:Rsemiinfexc}], or both [around both gaps of Fig.\ref{fig:Rsemiinfexcbragg}]. In the mixed exciton-Bragg regime, note that the reflectivity remains quite large in between the two gaps: the flatter the middle-polariton branch, the higher the corresponding reflectivity.
In the present semi-infinite geometry, reflectivity is complete inside the gaps where the wave vector becomes imaginary and the field inside the lattice consists of an evanescent wave.

The {\em output} problem corresponds to two counterpropagating Bloch modes with the same energy inside the lattice and a single transmitted plane wave in the external vacuum.
In this situation, the reflectivity is given by $R_{out}=|r_{out}|^2$ with
\begin{equation}
r_{out} = - \frac{E_+(l/2)+i(c/\omega)E'_+(l/2)}{E_-(l/2)+i(c/\omega)E'_-(l/2)},
\label{eq:Rout}
\end{equation}
where the $+$ and $-$ signs refer to the propagation versa of the polaritons. 

The system being invariant under time reversal and spatial parity, it is easy to prove that the coefficients $w_g$ and $y_g$ entering in the formula (\ref{eq:polaritonfield}) for polariton field share the same phase and can be chosen to be all real. The electric fields of counterpropagating polaritons are therefore complex conjugates of each other $E_-(x)=E_+(x)^*$. 
Plugging this into (\ref{eq:Rout}) and comparing the result with (\ref{eq:Rin}), it is immediate to see that $R_{in} = R_{out}$.

\subsubsection{Finite slab}

Reflectivity spectra are shown in Figs. \ref{fig:Rfinexc} and \ref{fig:Rfinexcbragg} for a finite system containing a quite large number $M$ of elementary cells. Note that the present approach is not able to calculate the reflectivity inside the gaps where the field consists of an evanescent wave.
The main difference with respect to the semi-infinite case is the presence of fast oscillations on top of the reflectivity spectrum around the main gaps. This can be explained as follows.

Two interfaces at respectively $x_{fr}=-((M-1)+1/2)l$ and $x_{back}=l/2$ now separate three regions of space: the vacuum with the incident and reflected photons, the finite-size lattice with counterpropagating polaritons, and again vacuum with now only a transmitted photon. 
The field in the last cell ($x\in[-l/2,l/2]$) is determined by the output problem to be
\begin{equation}
E_{st}(x) = E_+(x) + r_{out} E_-(x).
\end{equation}
As both $E_\pm(x)$ are Bloch states, the field in the first cell (taking $x\in[-((M-1)+1/2)l,-((M-1)-1/2)l]$) has the simple form
\begin{equation}
\tilde{E}_{st}(x) = E_+(x+(M-1)l)\,e^{-ikl(M-1)} + r_{out} E_-(x+(M-1)l)\,e^{ikl(M-1)}.
\label{eq:oscill}
\end{equation}
By solving the continuity conditions at the front interface at $x=x_{fr}$, we get
\begin{equation}
r_{slab} = e^{2i(\omega/c)x_{fr}} \frac{\tilde{E}_{st}(x_{fr}) + i(c/\omega) \tilde{E}'_{st}(x_{fr})}{\tilde{E}_{st}(x_{fr}) - i(c/\omega) \tilde{E}'_{st}(x_{fr})}.
\label{eq:Rfin}
\end{equation}
Because of the phase factors in (\ref{eq:oscill}), fast oscillations occur in the reflectivity (\ref{eq:Rfin}) due to the Fabry-Perot-like interference of Bloch waves which undergo multiple reflections at the lattice boundaries. 
The period $\Delta \omega$ of these oscillations is fixed by the group velocity $v^{gr}_{k,n}=d\Omega_{k,n}/dk$ and the total length of the system $L=Ml$,
\begin{equation}
 \Delta \omega= \frac{\pi}{L}\,v^{gr}_{k,n}:
\end{equation}
the slower $v^{gr}_{k,n}$, the closer the peaks.

To compare the envelope of this oscillations with the spectrum in the semi-infinite geometry, we can consider a simplified model where the lattice is replaced by a bulk medium of refractive index $n$.
In this case, the reflectivity for a single interface separating vacuum and medium is 
\begin{equation}
R_{int}=\left(\frac{1-n}{1+n}\right)^2.
\end{equation}
For a slab of thickness $L$, the reflectivity is~\cite{wait}
\begin{equation}
R_{slab}= \frac{(n - 1/n)^2\,\sin^2(\omega n L/c) }{4 \cos^2 (\omega n L/c) + (n + 1/n)^2 \sin^2(\omega n L/c)}
\end{equation}
Fabry-Perot oscillations are apparent, with a maximum reflectivity at the peaks equal to
\begin{equation}
R_{slab}^{max}=\left(\frac{1-n^2}{1+n^2}\right)^2.
\label{eq:Rslab}
\end{equation}
In the limit $n \rightarrow 1$, the ratio $(R_{slab}^{max}/R_{int}) \rightarrow 4$: this is due to the presence of two counterpropagating Bloch modes in the slab as compared to the single propagating mode in the semi-infinite case. 
This factor $4$ provides a good approximation in the lattice case as well, as one can easily see in the low-reflectivity tails of the spectra of Figs.\ref{fig:Rexc} and \ref{fig:Rexcbragg}.

The peak at Bragg frequency, which is more related to the photonic components, is generally smaller than the resonant one. This difference is dramatically enhanced in the purely excitonic regime as you can see in Fig.\ref{fig:Rfinexc}.

\section{1D lattice of three-level atoms}
\label{sec:1D3level}

Systems of three-level atoms are of great relevance for both quantum and non-linear optics. A most remarkable phenomenon in this respect is the so called \emph{Electromagnetically Induced Transparency} (EIT)~\cite{arimondo} which, among other properties, can lead to ultra-slow light propagation \cite{harris,matsko,hau} in spectral regions where absorption and reflectivity are also very low. This fact makes three-level systems extremely interesting systems to study and manipulate light in new regimes.

\begin{figure}[hbtp]
 \begin{center}
 \includegraphics[width = 0.4\textwidth]{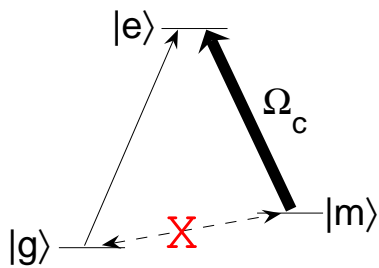}
 \caption{Sketch of the three-level $\Lambda$ configuration.}
 \label{fig:lambda}
 \end{center}
\end{figure}

We consider the three-level $\Lambda$ configuration shown in Fig.\ref{fig:lambda}: in addition to the previously considered two-level scheme, there is a metastable state $\left|m\right>$ which is coupled to the excited state by a classical (laser) field. The $\left|m\right> \Leftrightarrow \left|g\right>$ transition is optically forbidden.

The coupling beam dresses the metastable state, so to give a new mixed excitation formed by the atom promoted to the metastable state and an extra photon correspondingly emitted into the dressing beam: this new state has an energy $\hbar \tilde{\omega}_m = \hbar (\omega_m + \omega_c)$,  where $\omega_m$ and $\omega_c$ are the frequencies of respectively the bare metastable state (with respect to the ground state $\left|g\right>$) and the dressing laser light~\cite{CCT}.
Its detuning from the excited state is then $\delta_c = (\tilde{\omega}_m - \omega_0)$.
The Hamiltonian describing the dressing is
\begin{equation}
H_{c} =\frac{\hbar \Omega_c}{2} \sum_{i} \left(\hat{b}^{\dagger}_i \hat{d}_i + \hat{b}_i \hat{d}^{\dagger}_i \right),
\label{eq:coupling}
\end{equation}
where $\Omega_c$ is the Rabi frequency of the dressing beam and the operators $\hat{d}_i$, $\hat{d}^{\dagger}_i$ destroy and create the dressed metastable excitation for the atom $i$.

As above, the introduction of the radiation-matter interaction is performed by means of the minimal coupling replacement~\cite{ciuti4}:
\begin{equation}
\hat{b}_i \rightarrow \hat{b}_i - i\frac{\mu}{\hbar} \hat{A}(l_i).
\end{equation}
The $\hat{d}$ operators remain instead unchanged as the metastable state is not directly coupled to the quantized field.
After the minimal coupling replacement, the dressing Hamiltonian has the form:
\begin{equation}
H_{c} = \frac{\hbar \Omega_c}{2} \sum_{i} \left(\hat{b}^{\dagger}_i \hat{d}_i + \hat{b}_i \hat{d}^{\dagger}_i + i\frac{\mu}{\hbar}\hat{A}(l_i)\hat{d}_i - i\frac{\mu}{\hbar}\hat{A}(l_i)\hat{d}^{\dagger}_i \right).
\end{equation}
Note the appearance of terms directly coupling the dressed metastable state to the light field. These terms are however of the order of $m_c/\sqrt{\omega_0}$ (with $m_c= m\Omega_c/2\omega_0$), and therefore much smaller than all the other coupling strengths in the system.

\begin{figure}[hbtp]
 \begin{center}
 \includegraphics[width=\textwidth]{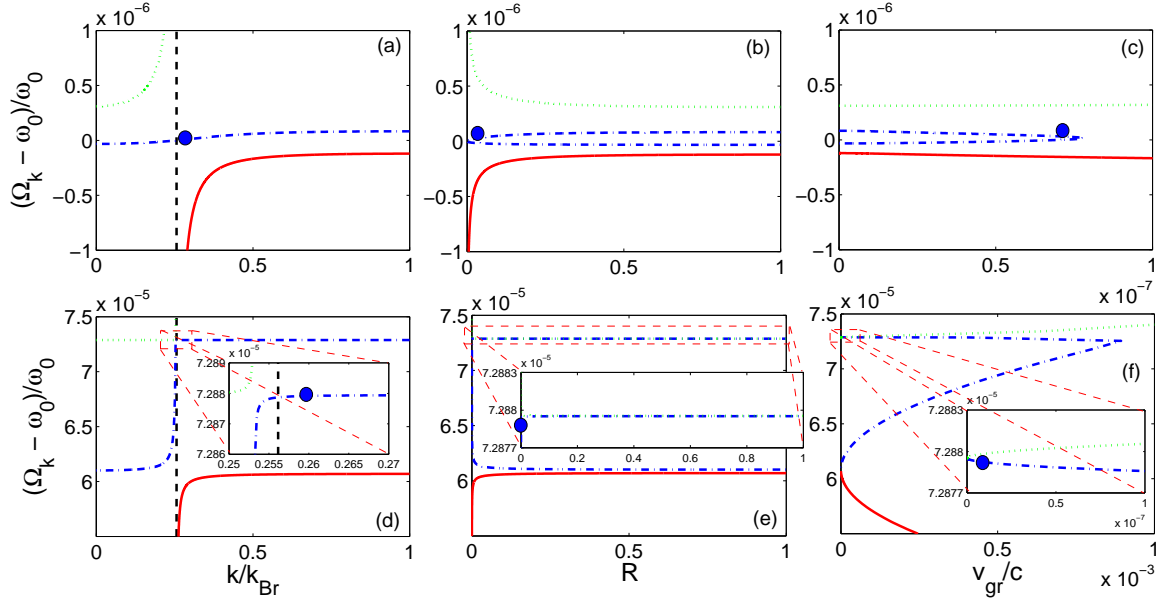}
\caption{Polariton properties in a 1D lattice of three-level atoms. Dispersion (a,d), reflectivity (b,e), group velocity (c,f) in the region around resonance. 
Red, solid line is lower polariton; blue, dot-dashed line is central polariton; green, dotted line is upper polariton; black, dashed line is photon dispersion in vacuum.
Parameters for a $l=100\,\textrm{nm}$ lattice of Rb atoms in the purely excitonic regime: $\omega_{Br} - \omega_0 \approx 3\omega_0$, exciton-photon coupling $m/\sqrt{\omega_0} \approx 4 \times 10^{-4} \hbar \omega_0$,  dressing amplitude $\Omega_c/\omega_0 = 2 \times 10^{-7}$.
Panels (a,b,c): $\omega \approx \tilde{\omega}_m = \omega_0$, $\delta_c = 0$ (initial state of the photon lifter). 
Panels (d,e,f): $\delta_c/\omega_0 = 1.2 \times 10^{-5}$ (final state of the lifter).
In the different panels, the blue circles indicate the position of the wavepacket to be ``lifted''.}
 \label{fig:lifter}
 \end{center}
\end{figure}

By following the same procedure as above, new excitonic delocalized operators can be constructed for the dressed state, and then included in the Bogoliubov matrix: the resulting polaritonic bands and the corresponding reflectivity spectra are shown in Figs. \ref{fig:lifter}(a,b) in the purely excitonic regime and for a resonant dressing $\delta_c=0$. Note that only the region around the atomic resonance is shown in the figure.

For the sake of simplicity, we concentrate our attention onto a purely excitonic case $\omega_0\ll\omega_{Br}$ where the lattice spacing is much smaller than the resonance optical wavelength.
The signature of the three-level scheme is the presence of a flat EIT mini-band corresponding to the ``central'' polariton. 
The most interesting region is around the \emph{Raman resonance} $\omega_k = \tilde{\omega}_m$ between the quantized radiation and the dressed state. This spectral region can be simply explored in terms of the reduced $3\times 3$ Bogoliubov matrix
\begin{equation}
\begin{pmatrix}
\hbar \delta_k & i m/\sqrt{\omega_0} & 0 \\
-im/\sqrt{\omega_0} & -\hbar \delta_c & \hbar\Omega_c/2 \\
0 & \hbar\Omega_c/2 & 0 
\end{pmatrix},
\label{eq:EITmatrix}
\end{equation}
where only the incident photon (1st row) mode, the excited state (2nd row), and the dressed metastable state (3rd row) are considered. $\delta_k=ck-\tilde{\omega}_m$ is the detuning of the photon mode from atomic resonance.
As we are considering a small region $\delta_k \ll \omega_0$, we have replaced $\omega_k$ with $\omega_0$ in the light-matter coupling terms. We have also neglected the $m'$ and $m_c$ coupling terms as they are much smaller than the other terms.

A noticeable property of (\ref{eq:EITmatrix}) exactly at Raman resonance $\delta_k=0$  is that for any value of different parameters there exists an eigenvalue $\lambda=0$ corresponding to the central polariton. Its eigenvector has a vanishing exciton component and a photonic weight
\begin{equation}
W_{ph} = \frac{\Omega_c^2}{\Omega_c^2+(4 m^2/\hbar^2\omega_0)},
\label{eq:Wph}
\end{equation}
so that the group velocity is 
\begin{equation}
v_{gr} = c\,\frac{\Omega_c^2}{\Omega_c^2+(4 m^2/\hbar^2\omega_0)}:
\label{eq:vgr}
\end{equation}
a significant reduction of the polariton group velocity is then observed as soon as a weak dressing amplitude is used $\hbar \Omega_c \ll m/\sqrt{\omega_0}$ (see Fig.\ref{fig:lifter}(c)).
Furthermore, reflection at the lattice interfaces is vanishing in the central region of the central polariton band, which allows for easy injection of polaritons in the system (as shown in Fig.\ref{fig:lifter}(b)).

A simple model to quantify the width of the region where reflection is low can be developed as follows. 
Defining an  effective index for the lattice as $n_{eff}= ck/\Omega_k=(c k_0 + \delta_k)/(c k_0 + \delta_k v_{gr}/c)$, and applying classical reflection result (\ref{eq:Rslab}), the reflectivity envelope for a semi-infinite system turns out to be approximately given by:
\begin{equation}
R_{int} \approx \frac{(\omega - \omega_0)^2}{4\,\omega^2_0}\frac{c^2}{v_{gr}^2}.
\label{eq:refdip}
\end{equation}
This approximation is accurate at the center of the dip where the reflectivity is small.
Comparing the two main results (\ref{eq:vgr}) and (\ref{eq:refdip}) of the present section, a trade-off is apparent: the slower the light, the narrower the reflectivity dip~\cite{matsko}.

\section{Application: photon energy lifter}

\label{sec:lifter}
Obtaining a coherent and widely tunable frequency conversion of an optical signal is a central task in optical telecommunications~\cite{freqconv}. Several techniques have been developed during the years to perform this operation, but most of them suffer from significant limitations in their application range, or are disturbed by spurious effects.

A recent proposal is based on the so-called dynamic photonic structures (for a review see e.g. ~\cite{dynphotstru}), whose optical properties are varied in real time while the optical wavepacket is propagating inside them.
The basic idea of the photon lifter consists in the adiabatic shift of the photonic band on which the photon is located.
This was orginally proposed for solid-state photonic structures~\cite{zeno}, but it is interesting to explore the potential of cold atom systems to this purpose: the very long coherence times in a Mott insulator state and the easy tunability by external electric or magnetic fields makes them very promising for this kind of applications.

As a specific example, we shall consider in what follows a system of $^{87}$Rb atoms, trapped in a cubic optical lattice of spacing $l = 100$nm. The optical properties are varied by means of an external magnetic field (taken as parallel to the $z$ axis) which shifts the atomic levels via the Zeeman effect~\cite{brehmmullin}.\\
In the following, we concentrate on the $D_2$ transition at a frequency $\omega_0/2\pi = 384\,\textrm{THz}$~\cite{dataline}.
As we are interested in substantial shifts, we can concentrate our attention in the high field regime ($B > 5\times 10^3\,\textrm{G}$) where the atomic nucleus is decoupled from the electronic degrees of freedom, and the energy shift mostly comes from the electronic total angular momentum only: $\Delta E = \mu_B g_J J_z B_z$, where $\mu_B$ is the Bohr magneton, $g_J$ is the Land\'e factor of the considered level and $J_z$ is the $z$ component of the total angular momentum of the electron.
We use the $|J=1/2,J_z=\mp 1/2\rangle$ sublevels of the $5^2S_{1/2}$ electronic ground state as respectively ground $\left|g\right>$ and metastable $\left|m\right>$ states, and the $|J=3/2,J_z=1/2\rangle$ sublevel of the $5^2P_{3/2}$ electronic excited state as excited $|e\rangle$ state. The corresponding Land\'e factors are $g_{J=1/2} = 2$ and $g_{J=3/2} = 4/3$.
The nucleus is not affected by the optical process and maintains the same polarization it had in the initial state: in the absolute atomic ground state, the nuclear spin is e.g. polarized antiparallel to the electron spin of the $|J=1/2,J_z=-1/2\rangle$ state.

A $z$ polarization is used for the dressing light beam that couples the $|m\rangle$ and the $|e\rangle$ states, and a $\sigma_+$ one is used to probe the polariton dispersion on the $|g\rangle \rightarrow |e\rangle$ transition. Using tabulated values for the electric dipole moment of the $D_2$ transition, the exciton-photon coupling (\ref{eq:exphcoupling}) for the system under consideration is of the order of $m/\sqrt{\omega_0} \approx 10^{-4} \hbar \omega_{0}$.

To maximize the available time to perform the lifter operation, it is useful to have a very slow group velocity, which in turns requires  a small dressing amplitude. In the following, we shall choose $\Omega_c/\omega_0=2\times 10^{-7}$. 
This value $\Omega_C/2\pi \approx 76\,\textrm{MHz}$ corresponds to $10$ times the radiative linewidth of the $D$ line of Rb atoms.

The dressing frequency is chosen in a way to have $\delta_c=0$ at the initial value $B_{in}$ of the magnetic field: the corresponding polariton dispersion is the one shown in Fig.\ref{fig:lifter}(a).
The light pulse is injected into the system in proximity of the resonant point $\delta_k=0$, where the interface reflectivity goes to zero, and injection is most effective (see the circle in Fig.\ref{fig:lifter}(a)): the width of this dip results from (\ref{eq:refdip}) to be of the order of $2\times 10^{-8} \omega_{0}$ and the group velocity (\ref{eq:vgr}) is $v_{gr}/c \approx 7 \times 10^{-8}$, i.e. $v_{gr} \approx 20\,\textrm{m/s}$.

The magnetic field variation is performed while the light pulse to be shifted is completely contained in the lattice and is propagating through an effectively bulk system.
As the magnetic field is varied in a spatially homogeneous way, the Bloch wave vector is conserved during the process.
If the field variation is slow enough as compared to the frequency difference of neighbouring bands, the polaritons will adiabatically follow the band and their frequency at the end of the process will be accordingly shifted (see the circle in Fig.\ref{fig:lifter}(d)).

As an example, we propose to tune the magnetic field from $1$ up to $2\,$T: this results in the metastable and excited states being shifted by respectively $(\delta_m - \delta_g)/\omega_0 = 7.3\times 10^{-5}$ and $(\delta_e - \delta_g)/\omega_0 = 6.1\times 10^{-5}$ with respect to the ground state. 
For light initially injected in proximity of $\omega_0=\tilde{\omega}_m$, the shift of the photon frequency results approximately equal to $\delta_m$, which amounts to the quite sizeable value $14\,\textrm{GHz}/\textrm{T}$.
As the lifter operation is based on an adiabatic shift of the polariton dispersion, it completely preserves the pulse shape and the coherence properties of the incident wavepacket, both at classical and at quantum level.

\section{Experimental issues}

\label{sec:experiment}

To verify the actual feasibility of such a promising experiment, it is important to mention the main practical difficulties that may arise in an actual experiment, and discuss how these can be overcome.

\begin{enumerate}
\item We have verified that the transmittivity of the lattice interfaces is close to $1$ for both the injection and the extraction process (Figs.\ref{fig:lifter}(b,e)).
The pulse is injected into the lattice at a frequency corresponding to the EIT reflectivity dip around Raman resonance.
The extraction takes place in close proximity of the Raman resonance where reflectivity is again very low. This, in spite of the fact we are very close to a gap: thanks to the now significant detuning $\delta_c$, the metastable state is in fact weakly coupled to light, and the corresponding crossing point is displaced slightly away from the light line.

\item In order to have a reasonably long time to vary the magnetic field, we have verified that the group velocity of the polariton states involved in the lifter operation is slow. Light initially propagates on the EIT slow light branch, which is deformed during the lifter operation. At the end, the wavepacket is found on the very flat region below the gap where the group velocity is low.

\item The wavepacket has to be shorter than the lattice length, still its frequency spectrum has to fit in the reflectivity dip at both injection and extraction. 
A lattice of $M=1000$ cells is able to accomodate pulses with at most $\Delta k \gtrsim 1/(lM)= k_{Br}/(\pi M)$. 
From panel (a), this corresponds to a lower bound on the frequency width of the incoming wavepacket $\Delta \omega_{in}=\Delta k\, v^{gr}_{in} > 5\times10^{-10}\, \omega_0$.
One can easily see in panel (b) that this frequency spread still fits within the injection window where reflectivity is low. 
The same can be verified on panels (d-e) for the extraction process.

\item In order for the pulse shape not to be affected, dispersion of the group velocity should be small for the wavevector window $\Delta k$ under examination. Initially, this is not a problem, as we are working close to the center of the EIT branch where the group velocity has a weak dispersion. The situation can be more critical on extraction, because of the strong squeezing of the polariton band in the region just below the gap.
The importance of this effect can be reduced by choosing pulses initially tuned just above the Raman resonance.
\end{enumerate}

One major constraint that still exists on the experimental parameters concerns the speed at which the magnetic field has to be actually varied. As this has to be done while the wavepacket is inside the lattice, a very slow group velocity and a long lattice are required.
Using values for state-of-the-art MIs. namely $L=M\,l=100\,\mu\textrm{m}$, and $v_{gr}=20\,\textrm{m}/\textrm{s}$, one obtains that one disposes of a time of approximately 
$5\,\mu\textrm{s}$ to perform the magnetic field variation. This means that a variation of $\Delta B=1\,\textrm{T}$ requires a very large rate of $2\,kG/\mu\textrm{s}$. 

As this can pose serious difficulties in an actual experiment, it is worth briefly exploring alternative strategies.
An interesting possibility is to further reduce the dressing amplitude $\Omega_c$.
As the polariton group velocity is proportional to the square of the dressing amplitude, the value $\Omega_c = 2 \times 10^{-8} \omega_0$ used in a recent slow light experiment~\cite{hau} already leads to a group velocity of the order of $20\; \textrm{cm/s}$ which corresponds to an available time of $500\,\mu\textrm{s}$.
In the high-field regime considered here, a photon frequency shift of $1\,\textrm{GHz}$ then requires a magnetic field variation of $500\,\textrm{G}$ in $500\,\mu\textrm{s}$, a rate routinely used in cold atom experiments.

It is important to note that the reduction of the dressing amplitude implies a squeezing of the reflectivity dip at injection and an enhancement of the dispersion at extraction. These, apparently serious problems are overcome thanks to the fact that a reduction in the group velocity implies a spatial shortening of the pulse in the lattice, and therefore a reduced frequency spread for a given length.

\section{Conclusions}
\label{sec:conclusion}

In conclusion, we have developed a fully quantum description of radiation-matter interaction in a gas of ultracold atoms trapped in the Mott insulator phase of an optical lattice. The coherent interaction between photons and atomic excitations gives rise to new, mixed polaritonic excitations.

In the case of two-level atoms, two different regimes are identified. In the purely excitonic regime, where the atomic resonance is far from the Bragg frequency of the lattice, two gaps appear in the energy spectrum. In the mixed exciton-Bragg regime, the interplay of the atomic resonance and the lattice periodicity enhances the gap amplitudes and gives rise to a flat band between them. The consequences of the polariton dispersion on the reflection properties of finite lattices have been investigated.

The theory is then extended to a system of three-level atoms in a $\Lambda$ configuration. The slow-light band which joins weak reflection to ultra-slow group velocity turns out to be the most promising in view of applications as a photon lifter, i.e. a device able to shift the carrier frequency of an optical pulse without affecting its shape nor spoiling its coherence properties. Advantages and disadvantages of using atomic gases as photon lifters are discussed.

Future work will address the application of slow as well as stopped polaritons as non-destructive probes of the microscopic properties of ultracold gases, e.g. the behaviour of matter wave coherence across the superfluid to Mott insulator transition.

\section{Acknowledgements}

We thank A. Recati, G. C. La Rocca, V. Guarrera and S. De Liberato for stimulating discussions. FB acknowledges hospitality at the  Institut Henri Poincare-Centre Emile Borel and financial support from ESF-QUDEDIS network through Short Visit Grant, Ref. Num. 1802. IC acknowledges hospitality at the  Institut Henri Poincare-Centre Emile Borel and financial support from CNRS.
FB dedicates his first paper in Physics to his family for the continuous and strong support received.

\bibliographystyle{jeos}

\end{document}